\documentclass[%
 reprint,longbibliography
showpacs,preprintnumbers,
bibnotes,
 amsmath,amssymb,
 prl,
]{revtex4-2}

\usepackage{graphicx}%
\usepackage[english]{babel}
\usepackage{microtype}          
\usepackage[breaklinks=true,colorlinks=true,linkcolor=blue,urlcolor=blue,citecolor=blue]{hyperref}

\usepackage{xcolor}     %
\usepackage{mathtools}
\usepackage{braket}
\usepackage{tikz}

 \usepackage{longtable}
  \usepackage{booktabs}
 \usepackage{siunitx}
 \usepackage{csquotes}
 
 \usepackage{multirow}
\usepackage{mathtools}
\usepackage{amssymb}

\newcommand{\refeqns}[2]{Eqs. (\ref{#1}) and (\ref{#2})}

\newcommand{\lowerbossphantom}{\vphantom{\bar{\bar{x}}}}
\newcommand{\upperbossphantom}{\vphantom{\dagger}}
\newcommand{\tempop}[3][\textstyle]{\settowidth{\dimen1}{$#1\hat{#2}$}\makebox[\dimen1][l]{$#1\hat{#2\mspace{#3}}$}}
\newcommand{\xop}[1]{{\mathchoice{\tempop[\displaystyle]{#1}{3.5mu}}{\tempop{#1}{3.5mu}}{\tempop[\scriptstyle]{#1}{3.5mu}}{\tempop[\scriptscriptstyle]{#1}{3mu}}}}
\newcommand{\chat}[1]{\ensuremath{\xop{#1}}}
\newcommand{\aop}[2]{\ensuremath{\chat{c}_{#1#2\lowerbossphantom}^{\upperbossphantom}}}
\newcommand{\cop}[2]{\ensuremath{\chat{c}_{#1#2\lowerbossphantom}^{\dagger\upperbossphantom}}}
\renewcommand{\i}{\mathrm{i}}
\renewcommand{\d}{\mathrm{d}}
\newcommand{\tn}[1]{\textnormal{#1}}

\usepackage{dcolumn}%
\usepackage{bm}%

\begin{document}
\preprint{APS/123-QED}

\title{Achieving the Scaling Limit for Nonequilibrium Green Functions Simulations}

\author{Niclas Schl\"unzen, Jan-Philip Joost, and
Michael Bonitz
 \email{bonitz@theo-physik.uni-kiel.de}}
\affiliation{
Institut f\"ur Theoretische Physik und Astrophysik, 
Christian-Albrechts-Universit\"{a}t zu Kiel, D-24098 Kiel, Germany
}

\date{\today}%

\begin{abstract}
The dynamics of strongly correlated fermions following an external excitation reveals extremely rich collective quantum effects. Examples are fermionic atoms in optical lattices, electrons in correlated materials, and dense quantum plasmas. Presently, the only quantum-dynamics approach that rigorously describes these processes in two and three dimensions is nonequilibrium Green functions (NEGF). However, NEGF simulations are computationally expensive due to their $T^3$ scaling with the simulation duration $T$. Recently, $T^2$ scaling was achieved with the generalized Kadanoff--Baym ansatz (GKBA), for second order Born (SOA) selfenergies, which has substantially extended the scope of NEGF simulations. Here we 
demonstrate that
GKBA-NEGF simulations can be performed with order $T^1$ scaling, both for SOA and $GW$ selfenergies, and point out the remarkable capabilities of this approach.   \end{abstract}

\maketitle
Strongly correlated fermion systems are attracting increasing interest in many fields including dense plasmas \cite{ebeling2017quantum, kremp-springer}, warm dense matter \cite{DORNHEIM_physrep18, gericke_cpp_19}, strongly correlated materials \cite{jensen_ultrafast_2013,joost_pss_18}, ultracold atoms \cite{schneider_fermionic_2012,schluenzen_prb16}, and atoms and molecules in strong radiation fields \cite{figueira_11_jmo,covito_epjb_18}. Of particular relevance are the relaxation phenomena that occur following an external excitation such as a rapid change (``quench'') of the confinement or the interaction strength, the impact of charged particles \cite{balzer_prb16,balzer_prl_18}, or the photoionization of atoms by lasers or free-electron lasers \cite{perfetto_pra_15,balzer_pra_10_2}.
Many theoretical approaches to the dynamics of strongly correlated fermions are limited either to one-dimensional systems (density-matrix renormalization-group simulations, DMRG) or short times (quantum Monte Carlo). The first quantum simulations of the expansion of correlated fermions in two and three dimensions were recently achieved using Nonequilibrium Green functions (NEGF) \cite{schluenzen_prb16} and exhibited very good agreement with experiments. The high accuracy of NEGF simulations was also demonstrated by comparison to DMRG \cite{schluenzen_prb17}.
However, these NEGF simulations are hampered by an unfavorable scaling with the simulation duration according to $T^3$ resulting from the two-time structure of the NEGF and the memory effects in the collision integral (see below).

This behavior can be relieved by applying the Generalized Kadanoff--Baym ansatz (GKBA) %
\cite{lipavski_prb_86,kalvova_pss_19}
which reduces the dynamics of the NEGF, $G(t,t')$, to propagation along the time diagonal, $t=t'$. It could be
demonstrated that, indeed, the expected improvement of the scaling, $N_\tn{t}^3 \to N_\tn{t}^2$ (in the following we will use the number of discretization time steps $N_\tn{t}=T/\Delta t$), can be achieved in practice for the  selfenergy in second order Born approximation (SOA) \cite{hermanns_jpcs13,balzer-book} where initial correlation effects can be treated even more efficiently \cite{karlsson_gkba18,bonitz_pss_18}.
 It could further be shown that this approximation, in many cases, does not lead to a loss of accuracy \cite{hermanns_prb14,schluenzen_prb17,covito_epjb_18}.
For these reasons, 
NEGF simulations using the GKBA with Hartree--Fock propagators (HF-GKBA) [cf. \refeqns{eq:gkba-def}{eq:occ-def} below] have become a powerful tool %
for studying the quantum dynamics in many fields, including optically excited semiconductors \cite{banyai_prl_98,lorke_jpcs_06, murakami_19}, excitonic insulators \cite{tuovinen_pss_19}, quantum transport and molecular junctions \cite{latini_charge_2014, kalvova_epl_18}, laser-excited plasmas \cite{kremp_99_pre,haberland_01_pre} and atoms \cite{perfetto_pra_15,covito_epjb_18}, strongly correlated electrons \cite{hermanns_prb14} and fermionic atoms in optical lattices \cite{schluenzen_cpp16,schluenzen_prb17}.
In recent years, significant effort was devoted to improve the GKBA, see, e.g., Refs.~\cite{latini_charge_2014,hermanns_prb14,karlsson_gkba18,verdozzi_gkba-note,kalvova_pss_19,tuovinen_efficient_2019,perfetto_dissection_2019,bonitz_pss_18}. %
Nevertheless, the quadratic scaling with $N_\tn{t}$ still makes the approach much less efficient than competing methods that scale linearly with $N_\tn{t}$, such as molecular dynamics, fluid theory, time-dependent density-functional theory within the adiabatic approximation or Boltzmann-type (Markovian) kinetic equations. 
In this Letter we demonstrate that 
the same linear scaling with $N_\tn{t}$, which is the ultimate limit in time-dependent simulations,
can be achieved for NEGF simulations within the HF-GKBA. This allows for unprecedented long simulations as well as for high-quality energy spectra  that are computed via Fourier transformation of time-dependent quantities, see, e.g., Refs.~\cite{kwong_prl_00, dahlen_prl_07, bonitz_cpp13}. We demonstrate this efficiency gain, compared to the original HF-GKBA, for finite Hubbard clusters and predict an even stronger gain for %
local selfenergies, such as for spatially homogeneous systems.
Moreover our approach allows one to compute additional quantities that are not directly accessible in standard NEGF schemes, such as the time-dependent pair-distribution function, the static and  dynamic structure factor, 
and various correlation functions.
Finally, we prove that linear scaling can be achieved also for more advanced selfenergies, such as $GW$, where all existing schemes scale as $N_t^3$.

\textbf{Theory.} We consider a general many-particle system with the Hamiltonian 
\begin{align}
    H(t) = \sum_{ij}h_{ij}(t)\cop{i}{} \aop{j}{} + \frac{1}{2}\sum_{ijkl} w_{ijkl} \cop{i}{} \cop{j}{} \aop{l}{} \aop{k}{},
    \label{eq:h-general}
\end{align}
containing a single-particle contribution, $\chat h$, and a pair interaction, $\chat w$. The matrix elements are computed with an orthonormal system of 
single-particle orbitals
$|i\rangle$.
The creation ($\chat c^\dagger_i$) and annihilation ($\chat c_i$) operators of particles in state $|i\rangle$ define the one-body nonequilibrium Green functions [correlation functions; here and below ``$\pm$'' refers to bosons/fermions], $G^<_{ij}(t,t')=\pm\frac{\i}{\hbar}\langle \chat{c}^\dagger_j(t')\chat{c}_i(t) \rangle$ and $ G^>_{ij}(t,t')=\frac{\i}{\hbar}\langle \chat{c}_i(t) \chat{c}^\dagger_j(t')\rangle$, where the averaging is performed with the correlated unperturbed density operator of the system.
The response of the system (\ref{eq:h-general}) to an external excitation is described by the Keldysh--Kadanoff--Baym equations (KBE) on the time diagonal \cite{lipavski_prb_86,balzer-book} where the Green function reduces to the single-particle density matrix, $\pm \i\hbar G^<_{ij}(t,t)=n_{ij}(t)$,
\begin{align}
    \frac{\partial n_{ij}(t)}{\partial t} - \frac{1}{\i\hbar}\sum_k [h^{\tn{HF}}_{ik}(t),n_{kj}(t)] = \pm I_{ij}(t)\,,
    \label{eq:kbe-diagonal}
\end{align}
with a mean-field Hamiltonian $h^{\tn{HF}}$. Here $I$ is the collision integral that takes into account interaction effects beyond Hartree--Fock, including scattering and dissipation, which we will treat in leading order, i.e., within the SOA
\cite{balzer-book,stefanucci_cambridge_2013}:
\begin{align}
\label{eq:colint}
 I_{ij}(t) 
 &= \left(\i\hbar\right)^2 \sum_{mnp} \, w_{imnp}\left(t\right) \sum_{kqrs} \int_{t_0}^t \d\bar t\, w^\pm_{qrsk}\left(\bar t\right) \times  \\
 \times &\left[ G^>_{nq}\left(t,\bar t\right) G^>_{pr}\left(t,\bar t\right) G^<_{sm}\left(\bar t,t\right) G^<_{kj}\left(\bar t,t\right) - (> \leftrightarrow <)
\right] \nonumber \, ,
\end{align}
where we defined
$ w^\pm_{qrsj} \equiv w_{qrsj} \pm w_{qrjs} =  \pm w^\pm_{qrjs}.$
Clearly, the computational effort to solve \refeqns{eq:kbe-diagonal}{eq:colint} scales with the number of time steps as $N_\tn{t}^2$. 

We now demonstrate that \refeqns{eq:kbe-diagonal}{eq:colint} can be reformulated such that the effort is reduced to $N_\tn{t}^1$ scaling. First, we introduce
an auxiliary four-index function $\mathcal{G}$,
\begin{align}
\label{eq:definition_g2}
 I_{ij}&(t) = \pm \i\hbar \sum_{mnp} w_{imnp}(t) \mathcal{G}_{npjm}(t) \,,\\
 \mathcal{G}_{npjm}&(t) = \i\hbar \sum_{kqrs} \int_{t_0}^t \d\bar t\, w^\pm_{qrsk}\left(\bar t\right) \times 
 \label{eq:g2-solution} \\
 \times &\left[ G^>_{nq}\left(t,\bar t\right) G^>_{pr}\left(t,\bar t\right) G^<_{sj}\left(\bar t,t\right) G^<_{km}\left(\bar t,t\right) - 
 (> \leftrightarrow <)\right]
\nonumber \, .
\end{align}
where 
the replacement $k\leftrightarrow s$ is used to match Eq.~(\ref{eq:colint}).
Comparing Eq.~(\ref{eq:definition_g2}) with the first equation of the Martin--Schwinger hierarchy for the many-particle Green functions \cite{martin_schwinger59} reveals that $\mathcal{G}(t)$ is nothing but the time-diagonal element of the two-particle Green function, %
and Eq.~(\ref{eq:g2-solution}) is its explicit form in second-Born approximation~\cite{bonitz_jpcs_13}.
Next, we introduce the GKBA \cite{lipavski_prb_86,balzer-book} (summation over $k$ is implied)
\begin{align}
    G_{ij}^\gtrless(t,t') &= 
    \pm G^{\rm R}_{ik}(t,t')n^\gtrless_{kj}(t') \mp 
    n^\gtrless_{ik}(t)G^{\rm A}_{kj}(t,t')\,,
    \label{eq:gkba-def}\\
    G^{\rm R/A}(t,t') &=\Theta[+/-(t-t')]\left\{G^\gtrless(t,t')-G^\lessgtr(t,t')\right\}\,,
    \nonumber\\
    \label{eq:occ-def}
n^<_{ij}(t) &= n_{ij}(t), \qquad n^>_{ij}(t) = n_{ij}(t) - \delta_{ij} \, ,
\end{align}
with Hartree--Fock propagators (``HF-GKBA''), $G^{\rm R/A} \to G^{\rm R/A, HF} $
and apply it to  each Green function in Eq.~(\ref{eq:g2-solution}):
\begin{eqnarray}
 \mathcal{G}^{\rm GKBA}_{npjm}(t) &=& \i\hbar \sum_{abcdkqrs}\int_{t_0}^t \mathrm{d}\overline{t}\, w^\pm_{qrsk}(\overline{t}) \times
 \label{eq:g2-solution-gkba}
 \\ && \nonumber \times
  \mathcal{U}_{npab}^{(2)}(t,\overline{t}) 
  \Phi^{absk}_{qrcd}(\overline{t})
  \mathcal{U}_{cdjm}^{(2)}(\overline{t},t)\, .
\end{eqnarray}
Here we introduced the abbreviations 
\begin{eqnarray}
\label{eq:phi}
 \Phi^{absk}_{qrcd}(t) &=&  \Phi^{absk>}_{qrcd}(t) -  \Phi^{absk<}_{qrcd}(t)\, ,
 \\
 \Phi^{absk\gtrless}_{qrcd}({t}) &=& %
n_{qa}^\gtrless({t})  n_{rb}^\gtrless({t}) n_{cs}^\lessgtr({t}) n_{dk}^\lessgtr({t})\,, \nonumber
\end{eqnarray}
and the two-particle time-evolution operator $\mathcal{U}^{(2)}$ is given in the supplementary material \cite{supplement}.

Finally, we remove the time integral in Eq.~(\ref{eq:g2-solution-gkba}) by differentiating  with respect to time which yields
\begin{align}
\label{eq:ode_g2}
\i\hbar \frac{\d}{\d t} & \mathcal{G}^{\rm GKBA}_{npjm} (t) - \left[h^{(2)\tn{HF}}, \mathcal{G}^{\rm GKBA}\right]_{npjm} (t) \\
 &= \frac{1}{\left(\i\hbar\right)^2}\sum_{kqrs} w^\pm_{qrsk}(t)
  \Phi^{npsk}_{qrjm}(t)  \, , \nonumber
\end{align}
where $h^{(2)\tn{HF}}_{ijkl}(t) = \delta_{jl} h^\tn{HF}_{ik}(t) + \delta_{ik}h^\tn{HF}_{jl}(t)$.
With this we have shown that NEGF theory within the HF-GKBA can be brought to a memory-less form (\ref{eq:ode_g2}) which, indeed, changes the scaling from quadratic to linear with $N_\tn{t}$. This was achieved by introducing the two-particle Green function on the time diagonal, $\mathcal{G}$, and by solving \textit{coupled time-local equations} for $G(t,t)$ and $\mathcal{G}(t)$. We, therefore, will refer to this as ``G1--G2'' scheme.
In fact, the one-to-one correspondence of NEGF theory within the HF-GKBA to time-local equations has been observed before \cite{hermanns_jpcs13,bonitz_qkt}. In Ref.~\cite{bonitz_qkt} it was also shown how to include arbitrary initial correlations, by supplementing Eq.~(\ref{eq:ode_g2}) with an initial value, $\mathcal{G}^{\rm GKBA}(t_0)=\mathcal{G}_0$. In Eq.~(\ref{eq:g2-solution-gkba})
this gives rise to an additional homogeneous solution
 that leads to an additional collision integral in the time-diagonal KBE (\ref{eq:kbe-diagonal}), in agreement with recent results~\cite{karlsson_gkba18,bonitz_pss_18}.
\begin{figure}[h]
\includegraphics[]{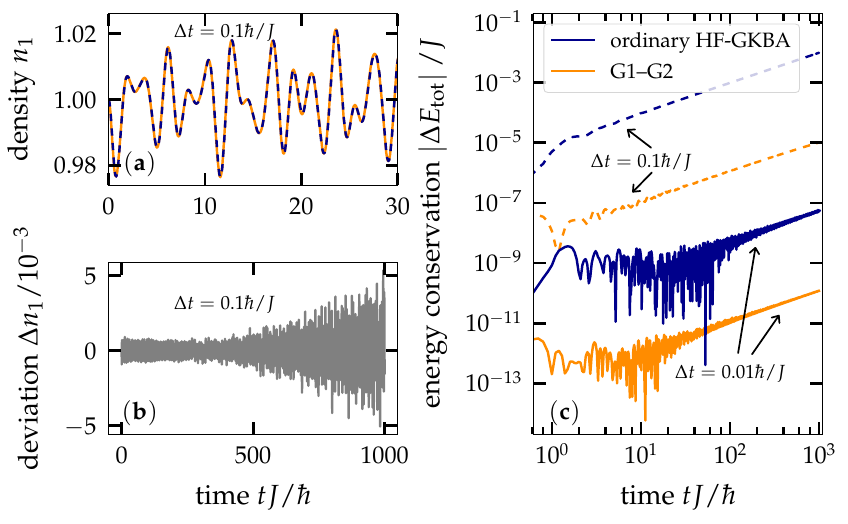}
\caption{\label{fig.energies} Comparison of the numerical accuracy of the ordinary HF-GKBA and the G1--G2 scheme with SOA selfenergies, for
 a four-site Hubbard chain with $U/J=1.5$, excited by a rapid potential change of amplitude $w_0 = 0.1J$ at site one. 
\textbf{(a)} Density evolution at the first lattice site, $n_1(t)$.
\textbf{(b)} Density difference between both methods, $\Delta n_1 (t)= n_1^\mathrm{G1-G2}(t) - n_1^\mathrm{ordinary}(t)$.
\textbf{(c)} Deviation from total-energy conservation for two time steps. Both NEGF implementations are based on a fourth-order integration scheme with the same time step with the initial state being prepared via adiabatic switching \cite{schluenzen_cpp16,hermanns_prb14}. }
\label{fig:densities}
\end{figure}

In the following we analyze the G1--G2 scheme more in detail. One readily confirms that the CPU time required to solve Eq.~(\ref{eq:ode_g2}),
\textit{for a general basis} of dimension $N_\tn{b}$, 
scales as $\mathcal{O}\left(N_\tn{b}^5 N_\tn{t}^1\right)$\footnote{The scaling reduction  $\mathcal{O}\left(N_\mathrm{b}^8\right) \to \mathcal{O}\left(N_\mathrm{b}^5\right)$ is achieved by reordering the occurring tensor contractions, cf. Eq.~(\ref{eq:phi}), \cite{schluenzen_jpcm_19}}.
In contrast, the original HF-GKBA scales as $\mathcal{O}\left(N_\tn{b}^5 N_\tn{t}^2\right)$ and, thus,
a dramatic speedup is expected, for any $N_\tn{b}$ 
\footnote{For comparison, the numerical effort of the above-mentioned initial-correlations scheme~\cite{karlsson_gkba18,bonitz_pss_18} scales as $\mathcal{O}\left(N_\tn{b}^5 N_{\tn{t}^*}^2\right)$ where $N_{\tn{t}^*}$ denotes the number of time steps after the build-up of correlations. 
On the other hand, a  scaling of the same order as our G1--G2 scheme has been achieved, e.g., in Ref.~\cite{perfetto_pra_15} by invoking the (drastic) Markov approximation to the GKBA equations, whereas the G1--G2 scheme is an identical reformulation of the original HF-GKBA.}.
At the same time, for many practical applications optimized basis sets are being used for which the scaling of both schemes has to be established separately. We, therefore, consider below two representative examples---the \textit{Hubbard} basis and the \textit{uniform electron gas}. The scaling for all three cases is summarized in  Tab.~\ref{tab:scaling}.

\begin{figure}[h]
\includegraphics[]{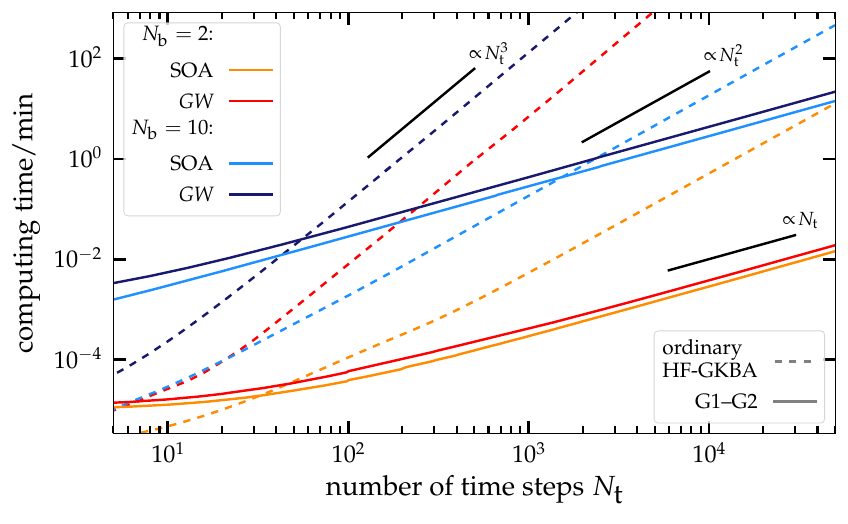}
\caption{\label{fig:time} Log-log comparison of the computational effort of the ordinary HF-GKBA (dashes) and the G1--G2 scheme (full lines) for Hubbard clusters 
as a function of propagation time $N_\tn{t}$. Colors denote system size and the selfenergy approximation.
}
\end{figure}

\textbf{Hubbard basis.}
The (Fermi--)Hubbard model is a key system in the theory of strongly correlated electrons in solids, see, e.g., Refs.~\cite{hubbard_1963,baeriswyl2013hubbard}, and it is being directly  realized with fermionic atoms in optical lattices, see, e.g., Refs.~\cite{esslinger_2010,schneider_fermionic_2012,TARRUELL_2018}. 
It is defined by the Hubbard Hamiltonian 
\begin{eqnarray}
 \chat{H}(t) = -J\sum_{\left<i,j\right>} \sum_\alpha \cop{i}{\alpha}\aop{j}{\alpha} + U(t) \sum_i \chat{n}_{i}^{\uparrow} \chat{n}_{i}^{\downarrow}\, ,
\label{eq:h-hubbard}
\end{eqnarray}
which includes hopping processes between nearest-neighbor sites $\left<i,j\right>$ with amplitude $J$ and an on-site interaction $U$, and $\alpha$ labels the spin projection. 
The integral (\ref{eq:definition_g2}) reads,
$I^{\uparrow (\downarrow)}_{ij}(t) = -\i\hbar U(t) \mathcal{G}^{\uparrow (\downarrow)}_{iiji}(t)\,, 
$
and the equation of motion (\ref{eq:ode_g2}) for $\mathcal{G}$ simplifies to
\footnote{The matrix $\mathcal{G}_{npjm}$ occurs with  two spin combinations ($\protect \uparrow$, $\protect \downarrow$) that are defined by the two possible combinations of spins on the r.h.s.
}
\begin{align}
& \i\hbar\frac{\d}{\d t} \mathcal{G}^{\uparrow (\downarrow)}_{npjm} (t) -  \left[h^{(2)\tn{HF}}_{\uparrow (\downarrow)}, \mathcal{G}^{\uparrow (\downarrow)}\right]_{npjm} (t) 
 = \left(\i\hbar\right)^2\sum_{k} U(t)\times
 \nonumber \\
   & \!\! \times \left[ G^{\uparrow (\downarrow) >}_{nk}(t,t) G^{\downarrow (\uparrow) >}_{pk}(t,t) G^{\uparrow (\downarrow) <}_{kj}(t,t) G^{\downarrow (\uparrow) <}_{km}(t,t) - > \leftrightarrow <\right]. \nonumber 
\end{align}
The numerical effort to solve this equation scales as $\mathcal{O}\left(N_\tn{b}^5 N_\tn{t}^1\right)$,
whereas the original HF-GKBA solution 
scales as $\mathcal{O}\left(N_\tn{b}^3 N_\tn{t}^2\right)$, cf. Tab.~\ref{tab:scaling}. It turns out that, for the Hubbard model, the new scheme exhibits the most unfavorable scaling with $N_\tn{b}$, as compared to the standard scheme and will become advantageous only for sufficiently large $N_\tn{t}$. For this reason we choose this case for numerical tests. In Fig.~\ref{fig:densities} we study the dynamics of a small Hubbard cluster and find excellent agreement between both schemes for all observables which is demonstrated for the density on site one in Fig.~\ref{fig:densities}.a,b. An even more sensitive accuracy test is the conservation of total energy. Here, the G1--G2 scheme turns out to be even more accurate than the standard HF-GKBA scheme if both use the same time step $\Delta t$, cf.~Fig.~\ref{fig:densities}.c.
We now compare in Fig.~\ref{fig:time} the CPU time required by both schemes for  Hubbard systems with $N_\tn{b}=2$ and $N_\tn{b}=10$. Our results clearly confirm the quadratic (linear) scaling with $N_\tn{t}$ of the original HF-GKBA (G1--G2) scheme as well as the predicted scaling with $N_\tn{b}$: when going from $N_\tn{b}=2$ to $N_\tn{b}=10$, ``break even'' is achieved for $N_\tn{t}$ approximately $(10/2)^2=25$ times larger, for SOA selfenergies.
\def\arraystretch{1.5}
\begin{table}
\begin{tabular}{l|c|c||c|c}
\multirow{2}{*}{\shortstack{Basis and\\ pair potential}}&\multicolumn{2}{c||}{SOA}&\multicolumn{2}{c}{$GW$}\\ \cline{2-3} \cline{4-5}
  & old & G1--G2 & old & G1--G2\\
 \hline
  general: $w_{ijkl}$ & $\mathcal{O}\left(N_\tn{b}^5 N_\tn{t}^2\right)$ & $\mathcal{O}\left(N_\tn{b}^5 N_\tn{t}^1\right)$ & $\mathcal{O}\left(N_\tn{b}^6 N_\tn{t}^3\right)$ & $\mathcal{O}\left(N_\tn{b}^6 N_\tn{t}^1\right)$ \\
 Hubbard: $U$ & $\mathcal{O}\left(N_\tn{b}^3 N_\tn{t}^2\right)$ & $\mathcal{O}\left(N_\tn{b}^5 N_\tn{t}^1\right)$ & $\mathcal{O}\left(N_\tn{b}^3 N_\tn{t}^3\right)$ & $\mathcal{O}\left(N_\tn{b}^5 N_\tn{t}^1\right)$ \\
  jellium: $v_{\left|\bm{q}\right|}$ & $\mathcal{O}\left(N_\tn{b}^3 N_\tn{t}^2\right)$ & $\mathcal{O}\left(N_\tn{b}^3 N_\tn{t}^1\right)$ & $\mathcal{O}\left(N_\tn{b}^3 N_\tn{t}^3\right)$ & $\mathcal{O}\left(N_\tn{b}^3 N_\tn{t}^1\right)$
\end{tabular}
\caption{\label{tab:scaling} Scaling of the CPU time with the number of time steps $N_\tn{t}$ and basis dimension $N_\tn{b}$ of the traditional non-Markovian HF-GKBA and the present time-local scheme (G1--G2), for three relevant basis sets, for SOA and $GW$ selfenergies.}
\end{table}

The \textbf{uniform electron gas (UEG, jellium)} is a key model for many-body physics, plasma and condensed-matter physics allowing to describe important features of the laser-driven nonequilibrium dynamics of electrons in metals \cite{rethfeld_prb_13}, warm dense matter \cite{gericke_cpp_19}, and quantum plasmas \cite{kremp_99_pre,kremp-springer}, as well as of electron--hole plasmas in semiconductors, see, e.g., Refs.~\cite{murakami_19,kwong-etal.98pss,binder-etal.97prb,haug_2008_quantum}.
Due to homogeneity, a momentum  (plain-wave) basis is advantageous where the Green function becomes diagonal: $G_{\bm{p}\bm{q}}\left(t,t'\right) 
:= \delta_{\bm{p}\bm{q}} G_{\bm{p}}\left(t,t'\right)$, for momenta $\bm{p},\bm{q}$. The Hamiltonian of the UEG in second quantization reads \cite{DORNHEIM_physrep18}
\begin{eqnarray}
 \chat{H}(t) &=& \sum_{\bm{p}\alpha} %
 \frac{\bm{p}^2}{2m}\cop{\bm{p}}{\alpha}\aop{\bm{p}}{\alpha} 
 + \sum_{\bm{p}\bm{p'}\bm{q}\alpha\beta} %
 v_{\left|\bm{q}\right|} \cop{\bm{p}+\bm{q}}{\alpha} \cop{\bm{p'}-\bm{q}}{\beta} \aop{\bm{p'}}{\beta} \aop{\bm{p}}{\alpha}\,,
\nonumber
\end{eqnarray}
with the Coulomb matrix element $v_{\left|\bm{q}\right|} = \frac{4\pi e^2}{\left|\bm{q}\right|^2}$. 
The integral (\ref{eq:definition_g2}) becomes
$I_{\bf{p},\sigma}(t) = \pm\i\hbar \sum_{\bf{\bar p},\bf{q},\alpha} v_{\left|\bf{q}\right|}(t) \mathcal{G}_{\bf{p} \bf{\bar p} \bf{q}}^{ \sigma \alpha }(t) \,,$
whereas the two-particle Green function $\mathcal{G}$ obeys
\begin{align}
 \i\hbar & \frac{\d}{\d t} \mathcal{G}_{\bf{p} \bf{\bar p} \bf{q}}^{ \sigma \alpha }(t) - \mathcal{G}_{\bf{p} \bf{\bar p} \bf{q}}^{ \sigma \alpha }(t) 
 \left(h^{\tn{HF}}_{\bm{p} -\bm{q},\sigma} + h^{\tn{HF}}_{\bm{\bar{p}} +\bm{q},\alpha} - h^{\tn{HF}}_{\bm{p},\sigma} - h^{\tn{HF}}_{\bm{\bar{p}},\alpha} \right)
 \nonumber
 \\\label{eq:g2eq-ueg}
 &= \left(\i\hbar\right)^2 
 \left\{v_{|\bm{q}|} (t) 
 \pm \delta_{\sigma,\alpha}
 v_{|\bm{p}-\bm{q}-\bm{\bar{p}}|} (t)\right\}
 \times 
 \\ 
  &\Big[G^>_{\bm{p} -\bm{q},\sigma}(t,t) G^>_{\bm{\bar{p}} +\bm{q},\alpha}(t,t) G^<_{\bm{p},\sigma}(t,t)  G^<_{\bm{\bar{p}},\alpha}(t,t) - (> \leftrightarrow <)\Big].
  \nonumber 
\end{align}
Interestingly, Eq.~(\ref{eq:g2eq-ueg}) scales as $\mathcal{O}\left(N_\tn{b}^3 N_\tn{t}^1\right)$ vs. $\mathcal{O}\left(N_\tn{b}^3 N_\tn{t}^2\right)$, for the standard HF-GKBA, cf. Tab.~\ref{tab:scaling}, and the G1--G2 scheme brings about a dramatic acceleration, independent of basis size.

\begin{figure}[h]
\includegraphics[]{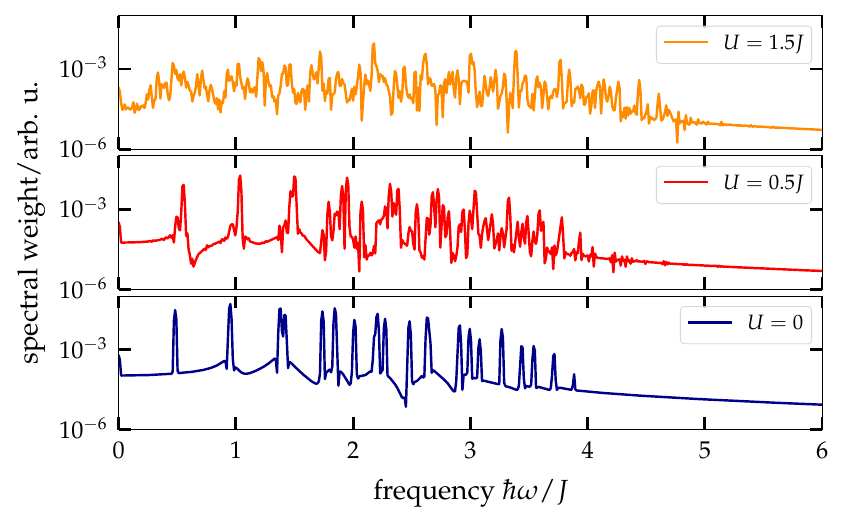}
\caption{
\label{fig:memory} 
Excitation spectrum of a 12-site Hubbard chain for three coupling strengths $U$ due to a rapid potential change of amplitude $w_0=0.01J$ at site one. The spectrum is obtained via Fourier transform of the density $n_1(t)$ computed with the G1--G2 scheme up to $T=600 \hbar/J$. The initial state is prepared using adiabatic switching.
}
\end{figure}

{\bf Spectra and two-particle observables}.
In addition to accelerating the time evolution, the G1--G2 scheme gives also access to more accurate spectral information. 
While, within the HF-GKBA spectral functions are treated on the Hartree--Fock level, correlation effects in energy spectra can be recovered by investigating the temporal response of the system to a short weak external excitation (linear response), see, e.g., Refs.~\cite{kwong_prl_00, dahlen_prl_07, bonitz_cpp13}.
This is demonstrated in Fig.~\ref{fig:memory} where the energy spectrum is retrieved via Fourier transform of the density evolution in a  Hubbard system.
Here the long propagation time enabled by the G1--G2 scheme allows us to resolve    correlation effects in the spectra, in particular broadening and shift of peaks as well as the emergence of new states at high energies. 

\begin{figure}[h]
\includegraphics[]{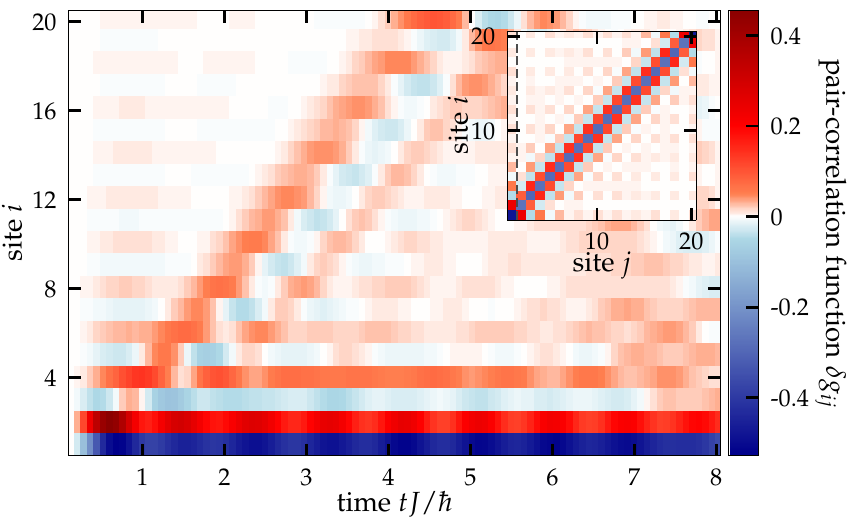}
\caption{\label{fig:pdf} Time evolution of the PCF relative to site 1, $\delta g_{i\uparrow,1\downarrow}(t)$, for a spatially homogeneous spin-symmetric 20-site Hubbard chain following an interaction quench $U/J=0\to 2$ at $t=0$. Sign-alternating correlations emerge rapidly and approach the correlated ground state (GS). Inset shows $\delta g^{\rm GS}_{i\uparrow,j\downarrow}$, for $U=2$, computed via adiabatic switching. Dashed line corresponds to data in the main figure.
}
\end{figure}

Furthermore, the G1--G2 scheme allows one to compute several quantities that are difficult or even impossible to access within standard NEGF schemes. This includes the nonequilibrium pair-distribution function (PDF) $g(\textbf{r}_1,\sigma_1;\textbf{r}_2,\sigma_2;t)$ and its Fourier transform---the static structure factor. 
Moreover, dynamic quantities, such as the density-- and spin-correlation functions or velocity-autocorrelation functions and the related transport coefficients---the dynamic structure factor, diffusion and absorption coefficients, and the dynamical conductivity within and beyond linear response---are becoming directly accessible.
In Fig.~\ref{fig:pdf}, we show, as an example, the time evolution of the pair-correlation function (PCF, i.e., the correlated part or the PDF) relative to site 1, $\delta g_{i\uparrow,1\downarrow}=g_{i\uparrow,1\downarrow}-n_{i\uparrow}n_{1\downarrow}$, for a 20-site Hubbard system after an interaction quench, $U/J=0 \to 2$. Initially the system is ideal, corresponding to  $\delta g\equiv 0$, and correlations emerge rapidly and spread with constant speed throughout the system. Changing $U$ does not affect this speed, but the amplitude of the distance-dependent oscillations is proportional to $U$.

\textbf{Extension to advanced selfenergies}. Finally, we test the G1--G2 scheme for the HF-GKBA with $GW$ selfenergies which are known to be significantly more accurate than SOA, in particular, at stronger coupling 
\cite{stefanucci_cambridge_2013,schluenzen_jpcm_19}. 
At the same time existing $GW$ simulations out of equilibrium scale as $N_{\rm t}^3$. Remarkably, we observe that the present G1--G2 scheme achieves order $N_{\rm t}^1$ scaling, as is summarized for different basis sets in Tab.~\ref{tab:scaling} (details are given in Ref.~\cite{supplement}, and similar results are observed for $T$-matrix selfenergies). The huge computational advantage brought about by G1--G2-$GW$ simulations becomes evident also in Fig.~\ref{fig:time}; even for Hubbard systems the G1--G2 scheme is advantageous, except for very short simulations. This indicates that a large class of problems is now becoming accessible for accurate NEGF simulations that had remained out of reach so far.

{\bf Summary and discussion}. We have implemented an alternative approach to NEGF simulations within the HF-GKBA that is memory-free and achieves the ultimate limit of linear scaling with the propagation duration $T$, as opposed to the common HF-GKBA approach with SOA ($GW$) selfenergies that is of order $T^2$ ($T^3$).
 This is achieved by 
 solving coupled time-local equations for $G(t,t)$ and the two-particle density matrix $\mathcal{G}(t)$. With this G1--G2 scheme we also established a direct link to the independent reduced-density-matrix (RDM) approach that has become popular in recent years in many fields, 
 see, e.g., Refs.~\cite{axt_zphysb_94, nakatsuji_prl_96, schmelcher, akbari_prb_12, bonitz_qkt,schuck_epja_16,lackner_pra_17}. Applying our derivation allows one to identify those RDM  approximations that are equivalent to common selfenergies in NEGF theory what enables one to make use of the full power of Feynman diagrams in RDM theory. 
We expect that the demonstrated astonishing scaling 
of the G1--G2 scheme will make highly accurate NEGF simulations of many nonequilibrium processes such as in laser-excited correlated systems achievable.

\section*{Acknowledgements}
We thank K. Balzer for valuable comments. This work was supported 
by grant shp00015 for CPU time at 
the Norddeutscher Verbund f\"ur Hoch- und H\"ochstleistungsrechnen (HLRN).

\bibliography{bibliography,dfg_pngf,mb-ref}

\end{document}